\documentclass[oupdraft]{bio}
\usepackage{url}
\usepackage{amsmath}
\usepackage{graphicx,psfrag,epsf}
\usepackage{enumerate}
\usepackage{natbib,comment}
\usepackage{amsfonts}
\usepackage{bbm}
\usepackage[hidelinks]{hyperref}
\usepackage{tabularx}
\usepackage{xcolor}

\begin{document}

\title{Inference for the Extended Functional Cox Model: A UK Biobank Case Study}

\author{ERJIA CUI$^1{}^\dag$, ANGELA ZHAO$^2{}^\dag{}^\ast$, ANDREW LEROUX$^3$, MARTIN A. LINDQUIST$^2$,\\
CIPRIAN M. CRAINICEANU$^2$\\[4pt]
\textit{1. Division of Biostatistics and Health Data Science, University of Minnesota, USA\\
2. Department of Biostatistics, Johns Hopkins Bloomberg School of Public Health, USA\\
3. Department of Biostatistics and Informatics, University of Colorado Asnchutz Medical Campus, USA}\\[2pt]
{azhao29@jh.edu}}

\footnotetext{These two authors contributed equally to this work.}

\markboth%
{E. Cui, A. Zhao, and others}
{Inference for Extended Functional Cox Models}

\maketitle
\footnotetext{To whom correspondence should be addressed.}

\begin{abstract}
{Multiple studies have shown that scalar summaries of objectively measured physical activity (PA) using accelerometers are the strongest  predictors of mortality, outperforming all traditional risk factors, including age, sex, body mass index (BMI), and smoking. Here we show that diurnal patterns of PA and their day-to-day variability provide additional information about mortality. To do that, we introduce a class of extended functional  Cox models and corresponding inferential tools designed to quantify the association between multiple functional and scalar predictors with time-to-event outcomes in large-scale (large $n$) high-dimensional (large $p$) datasets. Methods are applied to the UK Biobank study, which collected PA at every minute of the day for up to seven days, as well as time to mortality ($93{,}370$ participants with good quality accelerometry data and $931$ events).  Simulation studies show that methods perform well in realistic scenarios and scale up to studies an order of magnitude larger than the  UK Biobank accelerometry study. Establishing the feasibility and scalability of these methods for such complex and large data sets is a major milestone in applied Functional Data Analysis (FDA).}
{Functional data analysis; Survival analysis; Wearable devices; Biobank.}
\end{abstract}

\section{Introduction}
\label{sec1}
We introduce a class of extended functional Cox models and corresponding inferential tools designed to quantify the association between multiple functional and scalar predictors and time-to-event outcomes in large-scale (large $n$) high-dimensional (large $p$) datasets. We show that these models can be viewed as a particular case of mixed effects models, which provides a principled and scalable inferential framework. Methods are applied to the UK Biobank study, which collected physical activity (PA) at every minute of the day for up to seven days, time to mortality, as well as many other behavioral and demographic covariates ($93{,}370$ participants with good quality accelerometry data and $931$ events). Establishing the feasibility and scalability of these methods and providing ready to use, reproducible software that works on biobank-sized data sets is a major milestone in applied Functional Data Analysis (FDA).

\begin{figure}[h]
\centering
\includegraphics[width=16cm, height=16cm]{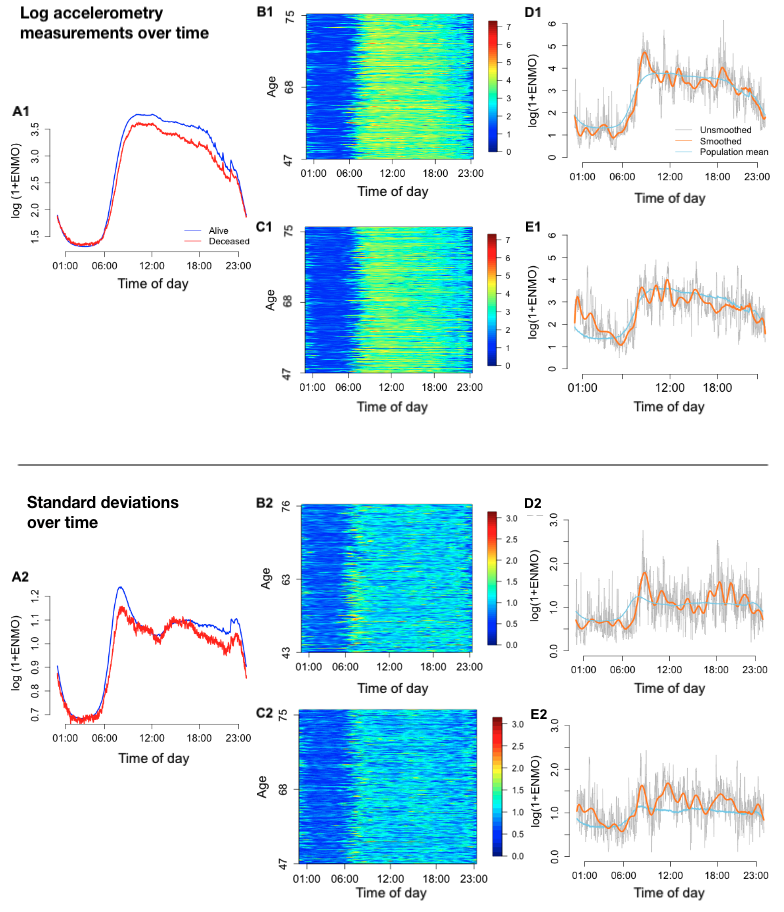}
\caption{Top five panels: minute-level log-transformed PA averaged across days within individuals. Left column: (A1) the mean log-transformed PA over time for individuals alive (blue) and deceased (red) as of July 31, 2020. Middle column: lasagna plots for minute-level log-transformed PA for 9310 alive (B1) and 931 deceased (C1) participants. Right column: smoothed (orange) and unsmoothed (gray) data of  one alive (D1) and one deceased (E1) participant overlayed with the population mean (cyan). Bottom panels display the same information as the top five panels but for the minute-level standard deviation across days within individuals. }\label{fig:PA_example}
\end{figure}

The emergence of biobank data sets has raised many computational and methodological challenges that range from handling very large data sets, accounting for non-rectangular data structures (e.g., a ``covariate" could be a scalar, such as age or BMI, or a high dimensional structure, such as minute-level PA data or images), to estimation and inference. Examples of such biobank studies are the UK Biobank \citep{allen2012ukbiobank}, the United States All of Us \citep{allofus2019}, and the China Kadoorie Biobank \cite{Chinabiobank}. Here we focus on the  UK Biobank, a large study conducted in the United Kingdom containing demographic, lifestyle, medical, and genetic information for around $500{,}000$ participants. 
In addition to this information, UK Biobank collected  high-quality accelerometry data for around $100{,}000$ participants, or about $20$\% of the original sample.

To illustrate the PA data structure, Figure~\ref{fig:PA_example} displays the minute-level PA data expressed as Euclidean Norm Minus One (ENMO, \cite{bakrania2016intensity}) separated by individuals who were still alive or deceased as of July 31, 2020 in the UK Biobank. For reference, PA accelerometry data were collected between 2013 and 2015. For the purpose of this paper we used PA data summarized at the minute level for the duration of the accelerometry substudy, which could be up to seven days for every study participant. 
The top panels display the minute-level means of the log-transformed PA, while the bottom panels display the minute-level standard deviations of the log-transformed PA. Both of these measures are calculated at a given minute of the day across days within each individual. The panels in the middle columns in Figure~\ref{fig:PA_example} display the PA data for one individual per row where participants are ordered by age, and data are separated by individuals who were still alive (panels B1 and B2) and deceased (panels C1 and C2) by July 31, 2020. The x-axis represents time in minutes from midnight to midnight and the y-axis is the $\log(1+{\rm ENMO})$ value, where ENMO is calculated at the minute level.  The left panels (A1 and A2) are the average across study participants within each group, where blue corresponds to study participants who were alive. More precisely, the blue line in panel A1 is obtained by taking the column averages in panel B1, while the red line in panel A1 is obtained by taking the column averages in panel C1. 
Together, these panels suggest that study participants who were alive had higher baseline mean activity and standard deviation between $7$ am and $11$ pm.   
The right columns provide examples of individual trajectories (gray), smooth estimators (orange), and corresponding population means (blue) for one participant who was alive (D1 and D2) and one who died (E1 and E2) before July 31, 2020. The panels D1 and E1 correspond to the mean, while D2 and E2 correspond to standard deviations across days at the same minute within the same person.

Using data from multiple studies \citep{agarwala2024quantifying, ledbetter2022cardiovascular,smirnova2020predictive, tabacu2020}, including UK Biobank \citep{leroux2021quantifying}, we have shown that summaries (e.g., mean, time in range) of the objectively measured minute level PA data displayed in Figure~\ref{fig:PA_example} are the strongest predictors of mortality, with better prediction performance than all traditional risk factors, including age, sex, body mass index (BMI), and smoking. The main goal of this paper is to investigate whether the complex, high-dimensional ($1{,}440$ observations per day), multivariate (minute-level mean and standard deviation at every minute over available days) PA data provides additional information about mortality, while accounting for other covariates. To do that we treat minute-level PA data as functional data \citep{crainiceanu2024functional,cui2023fast, kokoszka2017introduction,ramsay2005fitting}. Some of earliest methods for studying the association between functional data and time-to-event outcomes have been introduced by \cite{gellar2015cox} followed by several different methods \citep{cui2021additive, hao2021semiparametric,qu2016optimal, kong2018flcrm}; see Chapter 7 in \cite{crainiceanu2024functional} for a recent review of methods. The current paper expands on the existing literature in several directions by: (1) allowing for the inclusion of multiple functional, nonparametric, and parametric terms in the log hazard; (2) expanding inference by constructing pointwise and Correlation and Multiplicity Adjusted (CMA, \cite{crainiceanu2024functional}) confidence intervals and p-values; (3) scaling up methods to biobank-size data sets with over $90{,}000$ study participants and $1{,}440$ observations per study participant; (4) extending the models to incorporate distributions as predictors \citep{ghosal2023distributional2,ghosal2023distributional,ghosal2022scalar,petersen2022modeling} in a time-to-event model;  and (5) simulating for the first time survival models with high dimensional predictors of the size and complexity of the UK Biobank.
Finally, we provide reproducible software and discuss results of the UK Biobank data analysis. Such models, methods, and applications are crucial for increasing the relevance of functional data analysis methods in real world applications.

The paper is organized as follows.
We first introduce key components of statistical models and inferential tools in Section~\ref{sec:method}.
We then introduce the UK Biobank study, specific data processing steps, and the application results in Section~\ref{sec:ukb}.
Section~\ref{sec:sim} presents a simulation study.
We conclude with discussions in Section~\ref{sec:conc}.

\section{Methods}\label{sec:method}

\subsection{The Extended Functional Cox Models}\label{subsec:efcm}

For study participant $i = 1, \ldots, I$ we observe $Y_i = \min(T_i, C_i)$, where $T_i$ is the event time, $C_i$ is the censoring time, and  let $\Delta_i = \mathbbm{1}{(T_i \leq C_i)}$ be the binary event indicator. We assume that $C_i$ is independent of $T_i$ conditional on predictors, which is a realistic assumption in the UK Biobank where  censoring is primarily administrative. Denote by $\mathbf{Z}_i = [Z_{i1}, \ldots, Z_{ip}]^T \in \mathbb{R}^p$ the $p$ scalar predictors and by $\mathbf{X}_{i1}, \mathbf{X}_{i2}, \cdots, \mathbf{X}_{iR}$ the $R$ functional predictors, where $\mathbf{X}_{ir} = \{X_{ir}(s_r), s_r \in \mathcal{S}_r\}$ and $S_r$ could potentially be different domains of functional data. For the UK Biobank, $\mathbf{X}_{i1}$ and $\mathbf{X}_{i2}$ may represent the mean and standard deviation of the PA measure at the same minute of the day over multiple days, respectively. In this case $\mathcal{S}_1=\mathcal{S}_2$ is the set of time points expressed in minutes from midnight to midnight. In addition, denote by $\mathbf{V}_i = [V_{i1}, \cdots, V_{iq}]^T \in \mathbb{R}^q$ the $q$ scalar predictors assumed to have additive (nonlinear or linear) effects on the log hazard. An extended functional Cox model is
\begin{align}\label{eq:efcm}
    \log\lambda_i(t) = \log\lambda_0(t) + \mathbf{Z}_i^T\boldsymbol{\gamma} + \sum_{l = 1}^q f_l(V_{il}) + \sum_{r = 1}^R \int_{\mathcal{S}_r}{X_{ir}(s_r)\beta_r(s_r)ds_r}\;,
\end{align}
where $\lambda_i(t)=\lambda_i(t|\mathbf{Z}_i, \mathbf{V}_i, \mathbf{X}_{i1}, \cdots, \mathbf{X}_{iR})$ is the hazard function for the $i$th subject and $\lambda_0(\cdot)$ is the baseline hazard function. When $R = 1$ and $q = 0$, model~\eqref{eq:efcm} reduces to the linear functional Cox model proposed by \cite{gellar2015cox}. The functional coefficient $\beta_r(\cdot)$ quantifies the effect of the $r$th functional predictor at different locations of the functional domain. For example, if $X_{ir}(s_r)$ is the mean value of the PA intensity at time $s_r$ where $s_r = 1, \cdots, 1440$ is the time of day in minutes, $\beta_r(s_r)$ is interpreted as the effect of a one unit increase in PA intensity at time $s_r$ on the log hazard while keeping the rest of PA constant. In addition, $f_l(\cdot)$ represents an unknown, smooth function that captures the effect of the $l$th scalar predictor on the log-hazard of mortality. We anticipate that such models will become increasingly common not because they are complex, but because they provide practical answers to a range of scientific questions. They also naturally extend models for functional survival, semiparametric, and Generalized Additive Models (GAM) regression. Moreover, these methods have to adapt to the reality of large data sets, such as UK Biobank, which contains thousands of scalar and high-dimensional variables for hundreds of thousands of study participants.  

Model~\eqref{eq:efcm} is fit via basis expansion both for the additive and functional terms; for a general description of the approach see Chapters 3-7 in \cite{crainiceanu2024functional}. For $l = 1, \cdots, q$, denote by $\boldsymbol{\psi}_l(v) = [\psi_{l1}(v), \ldots, \psi_{K_{la}}(v)]^T$ the collection of $K_{la}$ spline functions defined over the range of the $l$th covariate, $V_{l}$. For $r = 1, \ldots, R$, denote by $\boldsymbol{\phi}_r(s_r) = [\phi_{r1}(s_r), \ldots, \phi_{K_{rb}}(s_r)]^T$ a collection of $K_{rb}$ spline functions defined over $\mathcal{S}_r$.  For $l = 1, \ldots, q$, $f_l(V_{il}) = \sum_{k=1}^{K_{la}} u_{lk} \psi_{lk}(V_{il})$ and for $r = 1, \ldots, R$, $\beta_r(s_r) = \sum_{k = 1}^{K_{rb}} b_{rk} \phi_{rk}(s_r)$, where $\mathbf{u}_l = [u_{l1}, \cdots, u_{lK_{la}}]^T$ and $\mathbf{b}_r = [b_{r1}, \cdots, b_{rK_{rb}}]^T$ are unknown spline coefficients. With these bases expansions, Model~\eqref{eq:efcm} becomes
\begin{align}\label{eq:efcm2}
\begin{split}
    \log\lambda_i(t) & = \log\lambda_0(t) + \mathbf{Z}_i^T\boldsymbol{\gamma} + \sum_{l = 1}^q f_l(V_{il}) + \sum_{r = 1}^R \int_{\mathcal{S}_r}{X_{ir}(s_r)\sum_{k = 1}^{K_{rb}} b_{rk} \phi_{rk}(s_r)ds_r} \\
    & = \log\lambda_0(t) + \mathbf{Z}_i^T\boldsymbol{\gamma} + \sum_{l = 1}^q \sum_{k=1}^{K_{la}} u_{lk} \psi_{lk}(V_{il}) + \sum_{r = 1}^R \sum_{k = 1}^{K_{rb}} b_{rk} \int_{\mathcal{S}_r}{X_{ir}(s_r)\phi_{rk}(s_r)ds_r} \\
    & = \log\lambda_0(t) + \mathbf{Z}_i^T\boldsymbol{\gamma} + \sum_{l = 1}^q \boldsymbol{\psi}_l(V_{il})^T \mathbf{u}_l + \sum_{r = 1}^R \mathbf{W}_{ir}^T\mathbf{b}_r\;,
\end{split}
\end{align}
where $\mathbf{W}_{ir} = [W_{ir1}, \cdots, W_{irK_{rb}}]^T$ and $W_{irk} = \int_{\mathcal{S}_r}{X_{ir}(s_r) \phi_{rk}(s_r)ds_r}$. This general structure is exactly the structure of the Cox regression model, though the covariates are designed specifically to handle functional data and nonparametric functions. Spline models are overparameterized and smoothing is induced by assumptions of the type $\mathbf{u}_l\sim N(\boldsymbol{0},\sigma^2_{ul}\boldsymbol{\Sigma}_{ul})$ and $\mathbf{b}_r\sim N(\boldsymbol{0},\sigma^2_{br}\boldsymbol{\Sigma}_{br})$, where $\boldsymbol{\Sigma}_{ul}$ and $\boldsymbol{\Sigma}_{br}$ are known and pre-specified for each specific spline basis, and $\sigma^2_{ul}$ and $\sigma^2_{br}$ are unknown variance parameters acting as smoothing parameters on the spline coefficients. The idea can be traced back to the papers of \cite{crainiceanugoldsmithwinbugs,goldsmith2011}, which first described it for scalar-on-function regression (SoFR); for a historical perspective and complete details, see \cite{crainiceanu2024functional}. These ideas were extended to time-to-event outcomes by maximizing the associated penalized partial log-likelihood \citep{crainiceanu2024functional,cui2023functional,leroux2020statistical, wood2017generalized}. 

While equation~\eqref{eq:efcm2} looks complicated, closer examination reveals that it can be viewed as a mixed effects model, where $\mathbf{u}_l$ and $\mathbf{b}_r$ are random effects parameters with a specific structure and $\boldsymbol{\gamma}$ are fixed effects parameters. Therefore, fitting model~\eqref{eq:efcm2} is equivalent to fitting a survival model with random effects, which can be done for some specific cases, but not in general. Indeed, in a recent monograph \cite{crainiceanu2024functional} it is shown how to fit related models using the {\ttfamily refund} \cite{refund} and {\ttfamily mgcv} \cite{wood2017generalized} \texttt{R} packages, though they were applied to a smaller data set (NHANES) \citep{leroux2019organizing}. We take this approach and show how to adapt it to the more complex models considered here. 

We now provide the \texttt{R} syntax to fit an extended functional Cox model when $p = 1$, $q = 1$, and $R = 2$. The formula can be extended to allow for more linear, additive, and functional terms. Suppose that the data are contained in the data frame \texttt{data} with the following fields: \texttt{time} is the observed time $Y_i$, \texttt{status} is the event indicator $\Delta_i$, \texttt{slVar} is a scalar predictor assumed to have linear effect on the log hazard, \texttt{saVar} is a scalar predictor assumed to have additive (nonlinear or linear) effect on the log hazard. In addition, \texttt{sFun1} is the matrix containing the first functional predictor where each row represents a subject and each column represents a location of the functional domain, and \texttt{sFun2} is the matrix of the second functional predictor stored in a similar format. For each functional predictor, we first create two matrices, \texttt{atmat} and \texttt{almat}, which are used to integrate functional observations over time.

\texttt{data\$almat1 <- I(matrix(1 / ncol(data\$sFun1), }

\hspace{35mm} \texttt{ncol = ncol(data\$sFun1), nrow = nrow(data)))}

\texttt{data\$atmat1 <- I(matrix(1:ncol(data\$sFun1), }

\hspace{35mm} \texttt{ncol = ncol(data\$sFun1, nrow = nrow(data), byrow = TRUE))}

\texttt{data\$almat2 <- I(matrix(1 / ncol(data\$sFun2), }

\hspace{35mm} \texttt{ncol = ncol(data\$sFun2), nrow = nrow(data)))}

\texttt{data\$atmat2 <- I(matrix(1:ncol(data\$sFun2), }

\hspace{35mm} \texttt{ncol = ncol(data\$sFun2, nrow = nrow(data), byrow = TRUE))}

\noindent After creating these matrices and storing them as four variables in \texttt{data}, the extended functional Cox model~\eqref{eq:efcm} can be fit using the \texttt{mgcv::gam} function as follows.

\texttt{fit\_efcm <- gam(time $\sim$ slVar + s(saVar) +} 

\hspace{33mm} \texttt{s(atmat1, by = almat1 * sFun1) + } 

\hspace{33mm} \texttt{s(atmat2, by = almat2 * sFun2), } 

\hspace{33mm} \texttt{data = data, weights = status, family = cox.ph())}

This approach to fitting complex functional Cox models is modular, as the additive term can be specified by applying the \texttt{s()} function to the variable, and including multiple functional predictors can be achieved by simply adding multiple \texttt{s(atmat, by = almat * sFun)} terms to the formula. Including additional nonparametric terms can similarly be done by adding terms of the type \texttt{s(sVar)}. Providing the model, describing the inferential approach, and introducing reproducible and extendable software implementation are important steps for promoting the use of functional data analysis methods in real data sets. 
The fact that these methods can be applied to data sets of the size and complexity of the UK Biobank data providesproof that the proposed methods are truly scalable and useful.

\subsection{Correlation and Multiplicity Adjusted Inference}\label{subsec:CMA}
 Consider the general case when $\beta(s) = \boldsymbol{\phi}(s)^T\mathbf{b}$ is one of the functional parameters and we obtain an estimator $\widehat{\mathbf{b}}$ of $\mathbf{b}$ and an estimator of its variance $\widehat{\mathbf{V}}_{\mathbf{b}}$  \citep{gray1992flexible,verweij1994penalized}. The pointwise $95$\% confidence interval for $\beta(s)$ can be constructed as $\widehat{\beta}(s) \pm 1.96\sqrt{\boldsymbol{\phi}(s)^T\widehat{\mathbf{V}}_{\mathbf{b}}\boldsymbol{\phi}(s)}$. Unfortunately, these intervals do not account for the multiple estimators or the complex correlations  along the functional domain. To address this problem we focus on correlation and multiplicity adjusted (CMA) confidence intervals, a nomenclature introduced in \cite{crainiceanu2024functional}. 

The idea of constructing confidence intervals taking into account the correlation along the domain was first introduced as ``simultaneous confidence bands" in the semiparametric regression setting by \cite{ruppert2003semiparametric}. It was later extended to the functional data setting by \cite{crainiceanu2012bootstrap}. Since then, the method was widely used in the literature \citep{cui2022fast,park2018simple, sergazinov2023case}. Here we use the label ``Correlation and Multiplicity Adjusted (CMA)" to explicitly refer to the inherent correlation of functional data and the many points where inference is conducted \cite{crainiceanu2024functional}. Let $\boldsymbol{\beta} = \{\beta(s), s \in \mathcal{S}\}$. Under the assumption that the bias of $\widehat{\boldsymbol{\beta}}$ is negligible and $\widehat{\mathbf{b}}$ is multivariate normal, we have $\widehat{\boldsymbol{\beta}} \sim \mathcal{N}(\boldsymbol{\beta}, \widehat{\mathbf{V}}_{\boldsymbol{\beta}})$, where $\widehat{\mathbf{V}}_{\boldsymbol{\beta}}(s, t) = \boldsymbol{\phi}(s)^T\widehat{\mathbf{V}}_{\mathbf{b}}\boldsymbol{\phi}(t)$. Denote by $\widehat{\mathbf{D}}_{\boldsymbol{\beta}} = \sqrt{\text{diag}(\widehat{\mathbf{V}}_{\boldsymbol{\beta}})}$ the square root of the diagonal of $\widehat{\mathbf{V}}_{\boldsymbol{\beta}}$. The goal of CMA inference is to find a critical value $q_{1 - \alpha}$ such that $P(|\widehat{\boldsymbol{\beta}} - \boldsymbol{\beta}| \leq q_{1 - \alpha}\widehat{\mathbf{D}}_{\boldsymbol{\beta}}) = 1 - \alpha$, where all inequalities are entrywise and simultaneous. In a recent monograph \citep{crainiceanu2024functional}, several methods were proposed to obtain $q_{1 - \alpha}$ in functional regression contexts. Here we use one of these approaches based on the assumption of  multivariate normality of $\widehat{\boldsymbol{b}}$, and hence $\widehat{\boldsymbol{\beta}}$. Notice that $q_{1 - \alpha}$ is essentially the equi-coordinate quantile of a multivariate normal distribution since $\widehat{\boldsymbol{\beta}}$ can only be observed on a finite grid. Therefore, given $\widehat{\mathbf{V}}_{\boldsymbol{\beta}}$, one can use existing software based on the multivariate normal distribution, such as the \texttt{mvtnorm::qmvnorm} function \citep{bornkamp2018calculating}, to calculate these quantiles. Suppose that $\widehat{\mathbf{V}}_{\boldsymbol{\beta}}$ is stored as a matrix \texttt{Vbeta}, the \texttt{R} syntax to obtain quantiles at level \texttt{alpha} is simply

\texttt{q <- qmvnorm(1 - alpha, corr = cov2cor(Vbeta), tail = "both.tails")\$quantile}

This approach is very fast, accounts for the correlations of the estimators, and can be implemented in one line of code. It has the advantage that it avoids resampling, which is much slower. In some cases, especially when the dimension of $\boldsymbol{\beta}$ is very large or the normality assumption of $\widehat{b}$ is suspect, it is a good idea to have an plan B. Suppose that $\widehat{\boldsymbol{\beta}}$ is evaluated on the grid $\{s_1, \cdots, s_L\}$, the critical value $q_{1 - \alpha}$ can be estimated as follows:
\begin{enumerate}
    \item Simulate $\beta^{(b)}(s_1), \cdots, \beta^{(b)}(s_L)$ from $\mathcal{N}(\widehat{\boldsymbol{\beta}}, \widehat{\mathbf{V}}_{\boldsymbol{\beta}})$;
    \item Calculate $r^{(b)} = \max_{l=1, \cdots, L}\{|\beta^{(b)}(s_l) - \widehat{\beta}(s_l)| / \widehat{\mathbf{D}}_{\boldsymbol{\beta}}(s_l) \}$;
    \item Repeat Step 1-2 for $b = 1, \ldots, B$ to obtain $r^{(1)}, \cdots, r^{(B)}$. 
    \item $q_{1 - \alpha}$ is estimated as the $(1 - \alpha)$ quantile of $r^{(1)}, \cdots, r^{(B)}$.
\end{enumerate}

After obtaining $q_{1 - \alpha}$ using either approach, the CMA confidence interval for $\beta(s)$ at level $\alpha$ is constructed as $\widehat{\beta}(s) \pm q_{1 - \alpha}\sqrt{\boldsymbol{\phi}^T(s)\widehat{\mathbf{V}}_{\mathbf{b}}\boldsymbol{\phi}(s)}$. Since the critical value is estimated as the quantile of the maximum of the quantiles along the functional domain, it is guaranteed to be larger than or equal to the quantile of a standard normal distribution. As a result, the CMA confidence intervals are uniformly non-narrower than the pointwise confidence intervals at a given significance level.

These CMA confidence intervals can be calculated for any value of $\alpha$. For every $s_l\in S$, we can then find the largest value of $\alpha$ for which the confidence interval does not include zero. We denote this probability by $p_{\rm pCMA}(s)$ and refer to it as the pointwise correlation and multiplicity adjusted  (pointwise CMA) p-value. As $q_{1-\alpha} \geq z_{1-\alpha/2}$, the pointwise CMA confidence intervals are non-narrower than the pointwise unadjusted confidence intervals. Therefore, the pointwise CMA p-values will be larger than or equal to the pointwise unadjusted p-values, leading to fewer or equal amounts of results being deemed ``statistically significantly different from zero." However, the tests will preserve the family-wise error rate (FWER) while accounting for test correlations.

Similarly, we define the global pointwise correlation and multiplicity adjusted  (global CMA) p-value as the largest $\alpha$ level at which at least one confidence interval $\widehat{\beta}(s_l)\pm q_{1-\alpha}\widehat{D}_{\beta}(s_l)$, for $l=1,\ldots, L$ does not contain zero. If we denote by $p_{\rm gCMA}(s_1,\ldots,s_L)$ this p-value, it can be shown that 
$$p_{\rm gCMA}(s_1,\ldots,s_L)=\min \{p_{\rm pCMA}(s_1),\ldots, p_{\rm pCMA}(s_L)\}\;$$
because the null hypothesis is rejected if it is rejected at any point in the domain of $\beta(s)$.

The advantage of using these p-values over the unadjusted p-values is that the tests preserve their nominal level. The pointwise tests focus on testing whether a particular value of the function is zero, whereas the global tests focus on testing whether the entire function is zero simultaneously.  

\subsection{Functional Distributional Cox Models}\label{subsec:dcm}

So far, we have focused on modeling the effects of functional predictors as a function of time, such as physical activity intensity at a particular time of the day, on time to event, such as mortality. An alternative approach is to consider the distribution of the time series of minute-level activities \cite{ghosal2023distributional,petersen2022modeling}. Here we extend the Cox regression models to quantify the association between the distribution of high-dimensional predictors and time to event. We show that techniques used for scalar-on-function regression can be extended to Cox models with distributional predictors.

Assume that  for the $i$th study participant, $i = 1, \ldots, N$, we transform the time series data $X_i(s)$ into the estimated quantile function, where $\widehat{Q}_i(p)$ is the $p$-quantile of the vector $\{X_i(s_1), \ldots, X_i(s_L)\}$. For the UK Biobank study, this quantile function provides information about the time spent at every level of physical activity, is easy to understand, and may be useful when comparing results from different studies. The functional distributional Cox model has the following form
\begin{align}\label{eq:dcm}
    \log\lambda_i(t|\mathbf{Z}_i, \mathbf{X}_i) = \log\lambda_0(t) + \mathbf{Z}_i^T\boldsymbol{\gamma} + \int_0^1{\widehat{Q}_i(p)\beta_Q(p)dp},
\end{align}
where $\beta_Q(p)$ quantifies the effect of the quantile corresponding to probability $p$ on survival outcomes. For functional models, the functional effects are incorporated by taking the integral over the functional domain. In the functional distributional Cox model, however, the effects are incorporated by taking the integral along the probability interval $[0, 1]$.  From here on, we propose to use exactly the same approaches described for functional Cox models. Moreover, these models can be extended  to the additive distributional Cox model \cite{cui2021additive}
\begin{align}\label{eq:adcm}
    \log\lambda_i(t|\mathbf{Z}_i, \mathbf{X}_i) = \log\lambda_0(t) + \mathbf{Z}_i^T\boldsymbol{\gamma} + \int_0^1{F\{p, \widehat{Q}_i(p)\}dp},
\end{align}
where $F(\cdot, \cdot)$ is an unknown bivariate twice differentiable function. This formulation allows a nonlinear association between quantile functions and hazard by introducing an additive structure. This model is beyond the scope of our current paper.

\section{UK Biobank Applications}\label{sec:ukb}

\subsection{Overview of UK Biobank}\label{subsec:overviewUK}

UK Biobank is a large prospective cohort study that includes $502{,}520$ participants registered with the National Health Service in England, Wales, and Scotland between 2006 and 2010 \citep{allen2012ukbiobank}. 
Comprehensive data on demographic, lifestyle, and health-related factors were collected. Extensive follow-up of participants' health and mortality status was conducted through linkage to national healthcare, death registries, and self-reports \citep{allen2012ukbiobank}. The UK Biobank received ethical approval (REC reference for UK Biobank 11/NW/0382) and participants have provided written informed consent. 

Among the $502{,}520$ participants, $236{,}519$ were invited to participate in an accelerometry substudy between 2013 and 2015. On average, individuals who were invited were younger and had better self-reported overall health than the rest of UK Biobank study participants \citep{leroux2021quantifying}. Of these participants, $103{,}700$ ($43.9$\% response rate) agreed to join the study. In this substudy, each individual was asked to wear an Axivity AX3 accelerometer on their dominant wrist for seven consecutive days immediately after receiving it. Accelerometry data was recorded continuously along three orthogonal axes at $100$ Hz with a dynamic range of $\pm8$g. After the monitoring period, participants were asked to mail-in the device back to the coordinating center using a pre-paid envelope \citep{doherty2017ukbaccel}.

\subsection{Data Processing}\label{subsec:dataprocessing}

Several quality control steps were performed after the raw accelerometry data was collected by the coordinating center. Among the $103{,}700$ participants, $7{,}164$ were identified as having poor data quality due to insufficient wearing time or incorrect device calibration. After excluding these participants, there  were $96{,}536$ study participants with ``good" quality accelerometry data, as determined by the UK Biobank team. The processed accelerometry data was released at multiple resolutions. In our analysis, we used the five-second average Euclidean Norm Minus One (ENMO, \cite{van2013separating}, expressed in milli-$g$’s ($g = 9.81 m/s^2$). The five-second data was then aggregated using 1-minute bins resulting in $1{,}440$ observations per day per study participant. To control data quality, we further excluded study participants with fewer than $3$ calendar days of accelerometry data and less than $1{,}368$ minutes of estimated wear time ($95$\% of the day.) The final data set consists of the minute-level accelerometry data from $93,370$ study participants, and is stored as a matrix with $1{,}440$ columns,
the so called ``1440 format" \citep{leroux2021quantifying}. To the best of our knowledge, this is the largest, best designed, and most quality-controlled study incorporating objectively measured PA to date.

In addition to the accelerometry data, multiple demographic, clinical, and health variables are available in the UK Biobank. The covariates included in our analysis are described in detail in Section~\ref{subsec:results}. The primary outcome in this paper is all-cause mortality, which was obtained from the UK Biobank data field 40000, ``Date of death". Individuals who died before the accelerometry measurement or have an invalid date of death were excluded from analysis. Mortality data available for this study includes observations up to June 13, 2020.

To reduce skeweness of the data, minute-level ENMO were transformed using the function $f(\text{ENMO}) =$ log(1 + ENMO) \citep{di2017patterns}. In some analyses the study-participant time series data were transformed into a distribution, where quantiles were obtained for each probability on a fine grid of probabilities between $[0,1]$.

\subsection{Results}\label{subsec:results}

We start by investigating the association between physical activity at different times of the day and the risk of all-cause mortality after adjusting for age, sex, BMI, overall health rating, smoking, drinking, diabetes, cancer, stroke, coronary heart disease, and Townsend deprivation index. An extended functional Cox model (labeled M1) was fit using our estimation and inferential procedure described in Section~\ref{sec:method}. We define the time of the accelerometry measurement as the baseline time. Age at the time of the accelerometry was obtained by subtracting the year of the measurement from the year of birth. To better understand how age affects log hazard, we modeled the effect of age nonparametrically. Sex, BMI, smoking, drinking, overall health rating, and Townsend deprivation index 
were obtained from their respective data fields. Diabetes, stroke, cancer, and coronary heart disease were stored as binary indicators of whether or not the participant was diagnosed with the condition at baseline.
Smoking and drinking indicators were assigned $1$ if the participant ever smoked or ever drank alcohol at the time of their initial assessment and $0$ if not. Participants with missing data for any of the above covariates were excluded (n = 850). These scalar covariates were added to our model as linear terms. PA data were further compressed by obtaining the mean and standard deviation of the minute-level log(1+ENMO) values across days at each minute $s$ of the day over all available days. For subject $i$, denote by $X_i(s)$ the mean of log(1+ENMO) at minute $s$, and by $U_i(s)$ the standard deviation of log(1+ENMO) at minute $s$ of the day.  With this notation our M1 model becomes
\begin{align}
    \log\lambda_i(t|\mathbf{Z}_i, \mathbf{X}_i) = \log\lambda_0(t) + \mathbf{Z}_i^T\boldsymbol{\gamma} + f_1(\text{age}_i) + \int_S{X_i(s)\beta_X(s)ds} + \int_S{U_i(s)\beta_U(s)ds},
\end{align}
where $f_1(\cdot)$, $\beta_X(\cdot)$, and $\beta_U(\cdot)$ are unknown smooth function that are estimated using penalized splines \citep{eilers1996flexible,ruppert2003semiparametric,wood2017generalized}. 

Suppose that \texttt{data} is a data frame organized as follows: \texttt{time} is the observed time $Y_i$, \texttt{status} is the event indicator $\Delta_i$ for mortality. The different scalar terms, $\mathbf{Z}_i$, are stored in their respective variable names. $X_i(s)$ and $U_i(s)$ are stored as two matrices of the same dimensions, \texttt{X} and \texttt{U}, where each row corresponds to $X_i(s)$ and $U_i(s)$, respectively. Similar to the general syntax introduced in Section~\ref{subsec:efcm}, we first create two matrices, \texttt{atmat} and \texttt{almat}, which are used to integrate the rows of \texttt{X} and \texttt{U} over time.

\texttt{data\$almat <- I(matrix(1 / ncol(data\$X), }

\hspace{33mm} \texttt{ncol = ncol(data\$X), nrow = nrow(data)))}

\texttt{data\$atmat <- I(matrix(1:ncol(data\$sFun), }

\hspace{33mm} \texttt{ncol = ncol(data\$sFun, nrow = nrow(data), byrow = TRUE))}

\noindent The model was fit using the \texttt{mgcv::gam} function as follows.

\texttt{M1 <- gam(time $\sim$ drinking + Townsend + cancer +}

\hspace{20mm} \texttt{smoking + chd + overall\_health + sex +}

\hspace{20mm} \texttt{s(age, k = 30, bs = "cr") +}

\hspace{20mm} \texttt{s(atmat, by = almat * X, bs = "cc", k = k) +}

\hspace{20mm} \texttt{s(atmat, by = almat * U, bs = "cc", k = k),}

\hspace{20mm} \texttt{data = data, weights = status, family = cox.ph())}

\begin{figure}[h!]
\centering
\includegraphics[width=14cm, height=18cm]{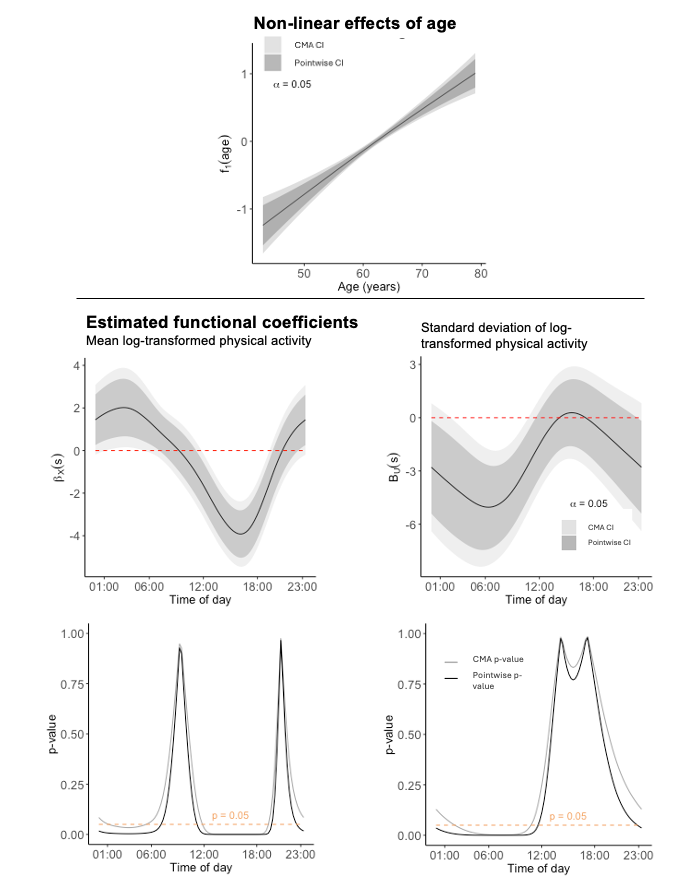}
\caption{Top panel: estimated effects of age, $f_1({\rm age})$, in the M1 model modeled using penalized splines. Second row: estimated functional coefficients, $\beta_X(s)$ (left panel) and $\beta_U(s)$ (right panel) together with $95$\% pointwise unadjusted (dark gray) and CMA (light gray) confidence intervals. The dashed red line corresponds to $0$. Third row: pointwise unadjusted (dark solid lines) and CMA (light gray lines) p-values corresponding to $\beta_X(s)$ (left panel) and $\beta_U(s)$ (right panel).
} \label{fig:res_M1}
\end{figure}

Figure~\ref{fig:res_M1} displays the estimated nonlinear effects of age, $f_1(s)$, functional coefficients $\beta_X(s)$ and $\beta_U(s)$, and their corresponding unadjusted and CMA confidence intervals and p-values across time of day. The $95$\% pointwise unadjusted and CMA confidence intervals are shown as dark gray and light gray regions, respectively. The CMA confidence intervals are uniformly wider than the pointwise unadjusted confidence intervals, which is expected, as discussed in Section~\ref{subsec:CMA}. Similarly, the CMA p-values are larger than the pointwise unadjusted p-values.

The top panel in Figure~\ref{fig:res_M1} displays the associations between age and log hazard of all-cause mortality. With increased age, the risk of mortality also increases and is statistically significant (p-value $< 0.0001$). This relationship appears linear and every one-year increase in age is associated with approximately $7.1\%$ increase in the hazard of mortality. 

The second row of panels in Figure~\ref{fig:res_M1} displays the point estimators of $\beta_X(s)$ and $\beta_U(s)$, which are the effects of the mean and standard deviation of physical activity at the time $s$ of the day. The dark gray areas correspond to the $95$\% pointwise unadjusted confidence intervals, while the light gray areas correspond to CMA confidence intervals. Based on the CMA confidence intervals for $\beta_X(s)$, we conclude that higher physical activity between 12pm and 7pm is significantly associated with a lower risk of all-cause mortality at $\alpha = 0.05$ even after accounting for other covariates and the standard deviation of physical activity. Note that this is slightly different from our previous results \citep{leroux2019organizing}, which indicate that physical activity between 8am and 8pm is significantly associated with lower risk of mortality. This is likely due to the fact that the current model accounts for the day-to-day variability, which is statistically significant exactly during the morning hours. These results are new and seem to indicate that higher day-to-day variability in the morning and higher afternoon physical activity are associated with lower risk of mortality. It is not immediately clear whether this provides new insights into ``what to do", but results may be due to the fact that physical activity is already relatively high in the morning for most individuals.   

The CMA and unadjusted p-values for $X_i(s)$ and $U_i(s)$ were obtained by inverting the respective confidence intervals and shown in Figure~\ref{fig:res_M1}. In addition to calculating a p-value for every time point, we also obtain the global p-value defined in Section~\ref{subsec:CMA}. For $X_i(s)$, this value was smaller than 0.0001 and for $U_i(s)$, this value was 0.0016.

For scalar predictors, we observed that in individuals 50 years and older, men at birth have $51.8$\% increased risk of mortality (p-value $< 0.0001$) compared to women. The effect of Townsend deprivation index is also significant (p-value = 0.0282) with an increase of one unit being associated with a $2.6$\% increase in the risk of mortality. For reference, the Townsend index in our data varies between -6.26 and 10.59, has a standard deviation of 2.81 with higher values corresponding to higher poverty. Likewise, past cancer diagnosis increases the risk of mortality  by $140$\% (p-value $< 0.0001$) and coronary heart disease increases it by $35.2$\% (p-value = $0.0034$). Participants that were previous and current smokers have $15.9$\% (p-value = $0.0449$) and $119$\% (p-value $< 0.0001$) increased risk of mortality compared to never smokers. Participants that have never drunk alcohol have $16.3$\% lower risk of mortality, though this was not statistically significant in our data set (p-value = $0.0668$) compared to those who drink. Finally, participant who self-reported overall significant health problems had a higher risk of mortality. Those who rated themselves to have excellent, good, and fair health have $45.1$\% (p-value = 0.0128), $41.3$\% (p-value = 0.0149), $31.4$\% (p-value = 0.0264) lower risk of mortality compared to individuals who rated themselves as having poor health status, respectively. The remaining scalar coefficients were not statistically significant.

\begin{figure}[h]
\centering
\includegraphics[width=8cm, height=8cm]{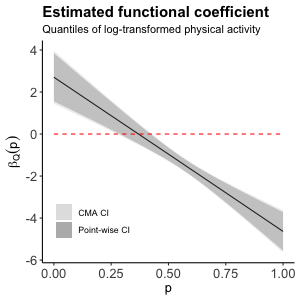}
\includegraphics[width=8cm, height=8cm]{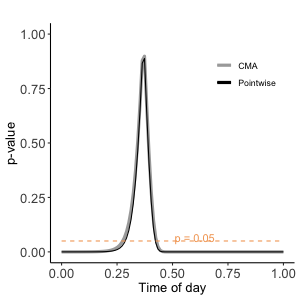}
\caption{Left panel: estimated functional coefficient $\beta_Q(p)$ from the M2 model as a function of the probability $p$. The model accounts for the list of covariates described in section \ref{subsec:results}. The $95$\% pointwise and CMA confidence intervals are shown by dark gray and light gray areas, respectively. A dotted red line is shown at 0. Right panel: Pointwise unadjusted and CMA p-values.}\label{fig:res_M2}
\end{figure}

Next, we applied the distributional Cox models introduced in Section~\ref{subsec:dcm} to explore the association between quantiles of physical activity and mortality. This model, labeled M2, adjusted for the same set of scalar predictors as model M1. Although the formulas look formally similar, model M2 provides different insights into how different intensity levels of physical activity are associated with the hazard of mortality. For model M2 we used a non-cyclic cubic penalized with 119 knots to model the $\beta_Q(s)$ functional coefficient as the effect is not expected to be similar at $p = 0$ and $p = 1$. For each individual, a subject-specific quantile function was estimated and used to fit the model. The estimation results are shown on the right panel of Figure~\ref{fig:res_M2}. The effect of quantile of physical activity is estimated to be linear in $p$, it is positive for $p \in [0, 0.38]$ and negative for $p \in (0.38, 1]$. We have not assumed linearity in the effect, so the result seems to suggest that $\beta_Q(p)$ could be safely assumed to be linear. We do not conduct here formal tests of linearity as they exceed the scope of the current paper.  Using CMA confidence intervals, we conclude that higher values of physical activity quantiles at $p = 0.45$ or higher and lower values of physical activity quantiles at $p = 0.3$ or lower are associated with lower risk of all-cause mortality (global p-value $< 0.0001$). A practical translation of these results is that being more active during the most active $10.8$ hours of the day ($0.55*24=13.2$) and being less active during the least active $7.2$ ($0.3*24$) hours of the day is associated with lower risk of all-cause mortality. Interestingly, one expects individuals to sleep between $6$ and $8$ hours every day, which is likely to be the period of lowest physical activity. This finding complements the results based on the analysis of the time series of physical activity. In particular, it suggests that restful sleep/inactivity and high physical activity during periods of activity are associated with reduced risk of mortality. We believe that this is the first time such results are reported using CMA inference.  

Similar to M1, several scalar predictors had statistically significant effects. Age (p-value $< 0.0001$) and Townsend deprivation index (p-value = $0.0158$) are significantly and positively associated with the risk of mortality. Having coronary heart disease (p-value = $0.0036$) or a biological male assignment at birth (p-value $< 0.0001$) also significantly increase the risk of mortality. Previous smokers (p-value = $0.0384$) and current smokers (p-value $< 0.0001$) are at significantly higher risk than never smokers. Individuals who self-reported as having excellent (p-value = $0.0109$), good (p-value = $0.0123$), or fair (p-value = $0.0199$) overall health have significantly lower risk of mortality compared to individuals who rated themselves as having poor health status, respectively.

Overall, these results highlight that the extended functional and distribution Cox models provide an efficient and interpretable solution to analyze the effects of physical activity on health outcomes in large-cohort studies. The applications of M1 and M2 reveal highly interpretable and complementary findings. The models described in this paper would not be computationally feasible without the rapid development of statistical software. Moreover, the syntax presented in our paper is reproducible and can be easily modified to include different scalar, additive, and functional terms depending on the problem setting. 

\section{Simulations}\label{sec:sim}

Simulations were conducted to: (1) provide a general framework for simulating functional predictors and survival outcomes from biobank-scale datasets; (2) evaluate the inferential properties of the proposed methods. For simplicity, we focus on the basic form of the model, with no scalar predictors and one functional predictor. 

\subsection{Synthetic Data Generation}\label{subsec:simdata}

We start by creating synthetic data sets that contain physical activity data and time-to-event outcomes that are as close as possible to the structure of UK Biobank. This is very difficult, as simulations from parametric survival distributions did not work due to their extreme sensitivity to small changes in parameters. To address this problem, we conducted simulations based on the estimated survival function obtained from fitting the extended functional Cox models to the UK Biobank data \cite{cui2021additive}. While these methods have been used in simpler models and smaller data sets, this was the first time that the extended functional Cox model was applied to a data set of the size and complexity of UK Biobank.  

\subsubsection{Simulating Functional Predictors}\label{subsec:simulatingfunctionalsurvival}

The first step was to simulate functional data that is similar to the physical activity data in the UK Biobank. Denote by $X_i(s)$ the subject-specific minute level average physical MIMS and let $\widehat{\mu}(s)$ be an estimator of the mean of $X_i(s)$. This could be obtained by taking a point-wise average, as the number of functions is very large, but one could also further smooth this average. Using Functional Principal Component Analysis (FPCA) we estimate the eigenfunctions, $\widehat{\phi}_k(s)$, and eigenvalues, $\widehat{\lambda}_k$, of the $X_i(s)-\widehat{\mu}(s)$ process.  This was achieved using the Fast Covariance Estimation (FACE) \citep{xiao2016} implemented in the {\ttfamily refund::fpca.face} function \cite{refund}. Physical activity data was then generated  from the model $\widehat{X}_i(s) = \widehat{\mu}(s) + \sum_{k=1}^M \xi_{ik} \widehat{\phi}_k(s) $, where $\xi_{ik} \sim \mathcal{N}(0, \widehat{\lambda}_k)$. Other distributions could be used, but here we use the normal distribution for simplicity. The non-negativity of the simulated data can be achieved by first applying the framework to log-transformed real data and then transforming the simulated data back to the original scale. For the purpose of this simulation we chose $M=35$, which ensured that 99\% of the variability was explained.

\subsubsection{Simulating Survival Outcomes}\label{subsubsec:sim_surv}
To simulate survival outcomes, we use the following framework described in \cite{cui2021additive}. This method requires a functional Cox regression model fit on the real data, the real survival outcome data, and the simulated functional predictors. First, the estimated cumulative baseline hazard, $\widehat{\Delta}_0(t) = \int_0^t{\widehat{\lambda}_0(\mu)du}$, was extracted from the fitted model. Next, the linear predictor $\widehat{\eta}_i = \int_\mathcal{S} \widehat{X}_i(s) \widehat{\beta}(s)ds$ was generated for each subject, where $\widehat{X}_i(s)$ was simulated as described in Section~\ref{subsec:simulatingfunctionalsurvival} and $\widehat{\beta}(s)$ was obtained from the fitted model. After obtaining $\widehat{\Delta}_0(t)$ and $\widehat{\eta}_i$, the estimated survival function is calculated as $\widehat{S}_i(t) = \exp\{-e^{\widehat{\eta}_i}\widehat{\Delta}_0(t)\}$. The survival time, $\tilde{T}_i$, is then simulated from a random variable with cumulative distribution function $1-\widehat{S}_i(t)$.  The censoring time $\tilde{C}_i$ was generated by sampling with replacement from the censoring times in the UK Biobank. The simulated survival outcomes are obtained as $\tilde{Y}_i = \min(\tilde{T}_i, \tilde{C}_i)$ and $\tilde{\Delta}_i = \mathbbm{1}{(\tilde{T}_i \leq \tilde{C}_i)}$, respectively.

\subsection{Estimation Accuracy}

The aim in this section is to evaluate how close are the results based on the simulated data to the results based on the real UK Biobank data. To do this, we fit a linear functional Cox model to the simulated data and compare the functional coefficient, $\beta(s)$, to the true coefficient. Here the true coefficient, denoted as $\widehat{\beta}_{\rm UKB}(s)$, is the coefficient obtained from the real UK Biobank data where the total sample size was N = $93{,}370$ and the number of events was $931$. We will simulate data of various total sample sizes, while maintaining the same proportion of events (about 1\%). 

We conducted $200$ simulations for each of the following sample sizes: $N = 50{,}000$, $100{,}000$, $500{,}000$, and $1{,}000{,}000$.  All simulations were conducted conditional on $\widehat{\beta}_{\rm UKB}(s)$, which is considered to be the ``true functional coefficient". Running such simulations is non-trivial, especially for larger sample sizes. For example, the $200$ simulations for the M1 model with $N=1{,}000{,}000$ observations required up to $400$GB of memory and $2.5$ days (about $18$ minutes per data set) on a cluster computing system. The code for our simulation study is available in our Github repository.

\begin{figure}[h]
\centering
\includegraphics[width=12cm, height=7cm]{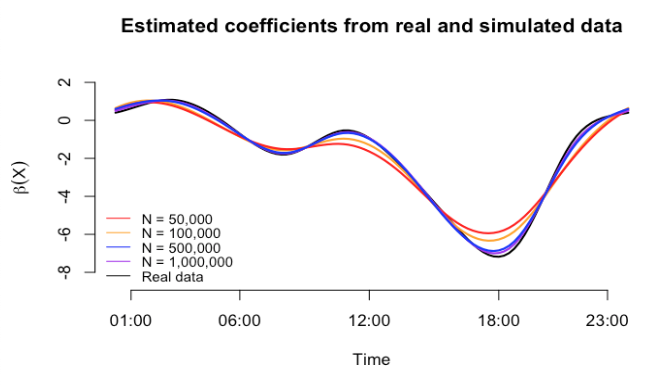}
\caption{The black line represents the estimated functional coefficient from a model fitted on the real UK Biobank data. The functional coefficient from models fitted on simulated data of different sample sizes from 200 simulations are plotted in color: N = 50,000 (red line), 100,000 (orange line), 500,000 (blue line), and 1,000,000 (purple line).}\label{fig:simulations}
\end{figure}

\begin{figure}[h]
\centering
\includegraphics[width=12cm]{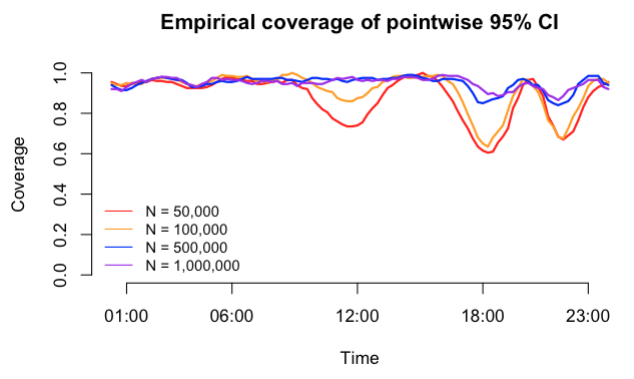}
\caption{The empirical coverage of the pointwise 95\% confidence intervals at each location of the functional domain under different sample sizes from 200 simulations.}\label{fig:pointwiseCI}
\end{figure}

The average estimated coefficients from simulated data with different sample sizes are shown in different colors in Figure~\ref{fig:simulations}. The true functional coefficient, $\widehat{\beta}_{\rm UKB}(s)$, is shown as the black line. The estimated average coefficients from simulated data are close to the true coefficient and get closer as sample size increases. This suggests that estimators have small bias and the bias gets smaller with increased sample size. To further quantify the estimation accuracy of simulated data, the mean squared error (MSE) and the empirical coverage of the $95$\% pointwise and CMA confidence intervals were calculated for each sample size and the results are shown in Table~\ref{tab:mse}.
As the sample size increases, the MSE decreases. The MSE value is $0.004$ for $N = 1{,}000{,}000$, compared to $0.291$ for $N=50{,}000$, or less than $2\%$. Furthermore, as the sample size increases, the empirical coverage of both of the pointwise and CMA confidence intervals increases. However, they are both below the nominal level for $N=100{,}000$, which is the approximate size of the original sample.  
Figure~\ref{fig:pointwiseCI} displays the empirical coverage of the pointwise 95\% confidence intervals along the functional domain. For smaller sample sizes, under coverage was observed at several locations of the domain where there is a change in the curvature. Thus, it is likely that under coverage may be partially attributed to biases in the point estimates in those regions, though  non-normality of estimators could also be a factor. Note that CMA confidence intervals have lower coverage than the average coverage of pointwise confidence intervals. This is expected because average pointwise confidence intervals is the average over all time points of the proportion of simulations where the pointwise confidence intervals cover $\widehat{\beta}_{\rm UKB}(s)$. In contrast, the coverage of CMA confidence intervals is defined as the proportion of times out of $200$ simulations when the confidence interval covers $\widehat{\beta}_{\rm UKB}$ at all time points. Thus, the CMA performance is more sensitive to bias anywhere along the time range. In fact this is more comparable to the minimum coverage of the pointwise estimators than the average; see Figure~\ref{fig:pointwiseCI}. 

\begin{table}
    \caption{The mean squared error and mean confidence interval (CI) coverage of the estimated functional coefficient based on 200 simulations with different sample sizes and proportion of cases.}
    \label{tab:mse}
    \begin{tabularx}{\textwidth} { 
     |>{\raggedright\arraybackslash}X 
    |>{\raggedright\arraybackslash}X 
    |>{\raggedright\arraybackslash}X 
    |>{\raggedright\arraybackslash}X 
    |>{\raggedright\arraybackslash}X | }
     \hline
     \textbf{Sample Size, N} &  \textbf{Mean Squared Error} & \textbf{Pointwise CI Coverage} & \textbf{CMA CI  Coverage}\\
     \hline
     50,000  & 0.291 & 87.7\% & 74.5\%\\
     \hline
    100,000 & 0.312 & 91.3\% & 81.5\%\\
    \hline
    500,000 & 0.006 & 94.8\% & 93.5\%\\
    \hline
     1,000,000  & 0.004 & 95.0\% & 92.0\%\\
    \hline
    \end{tabularx} 
\end{table}

\section{Discussion}
\label{sec:conc}
Inferential methods for survival analysis were extended to incorporate multiple functional predictors, nonparametric effects of scalar covariates, and parametric effects. For the first time, correlation and multiplicity adjusted (CMA) confidence bands and p-values for the functional coefficients were calculated. Distributional predictors (where the functional data are replaced by the quantile function of the observed values) were also incorporated and treated as a standard scalar-on-function regression model. Models were applied to the UK Biobank data set, which is one order of magnitude larger than previous data sets used for functional survival models \cite{crainiceanu2024functional}. The associated software is easy to use and scalable as it takes around $20$ minutes to fit the model to the UK Biobank. For reproducibility purposes, the associated {\ttfamily R} code for our analysis is available at \href{https://github.com/ang-zhao/biobank\_functional\_cox\_mortality}{https://github.com/ang-zhao/biobank\_functional\_cox\_mortality}.

An ever larger body of evidence indicates that summaries of objectively measured physical activity are the strongest predictor of mortality and many diseases in people over $50$ \cite{cui2022semiparametric,ledbetter2022cardiovascular,leroux2021quantifying,meng2023quantifying,smirnova2020predictive,  zhao2023evaluating}. Thus, there is a need to investigate whether the daily patterns of PA  contain more information than these summaries. This paper validates some of the published scientific results in a much larger data set and provides new scientific insights: (1) increased physical activity during the day and reduced physical activity during the night are associated with lower risk of death; (2) higher day to day variability in the morning conditional on the same level of activity in the morning is associated with lower mortality risk; (3) conditional on the functional effects of the mean and standard deviation of physical activity, the effect of age on the log hazard is indistinguishable from being linear; (4) when distributional predictors are used, the functional effect is estimated to be linear, positive below quantile $p=0.30$ (consistent with positive effects of less physical activity during rest periods) and negative above quantile $p=0.55$ (consistent with positive effects of more physical activity during active periods); and (5) even with very large studies, estimating and quantifying the inferential properties of estimators  is important, especially when the number of outcomes is small (e.g., $1$\% of the total sample size).

Another important contribution of the paper is to show how to simulate survival data with functional predictors that closely mimic the UK Biobank data. Simulations indicate that inference results are close to their nominal level for sample sizes close to the UK Biobank and estimation and coverage probabilities continue to improve with increased sample size at the same percentage of observed outcomes (1\%). 

Functional regression models, and applied functional data analysis more generally, like other statistical methods, will continue to be successful if it can adapt to larger and more complex data, provide insights that may not be available from data summaries, or suggest simpler models without substantial loss of information. Here we propose a set of analytic tools that satisfies these requirements, which provides a necessary step forward in the analysis of high-dimensional functional data.

\bibliographystyle{biorefs}
\bibliography{ref} 

@article{agarwala2024quantifying,
  title={Quantifying the time-varying association between objectively measured physical activity and mortality in US older adults over a 12-year follow-up period: the {NHANES} 2003--2006 study},
  author={Agarwala, N. and Zipunnikov, V. and Crainiceanu, C.M. and Leroux, A.},
  journal={BMJ Evidence-Based Medicine},
  year={2024},
  publisher={Royal Society of Medicine}
}

@article{allen2012ukbiobank,
    title={{UK} Biobank: Current status and what it means for epidemiology.},
    author={N. Allen  and C. Sudlow and P. Downey and T. Peakman and J. Danesh  and P. Elliott and J. Gallacher and J. Green and P. Matthews and J. Pell and T. Sprosen and R. Collins  on behalf of UK Biobank},
    journal={Health Policy and Technology},
    pages={123--126},
    year={2012}
}

@article{bakrania2016intensity,
  title={Intensity thresholds on raw acceleration data: Euclidean norm minus one (ENMO) and mean amplitude deviation (MAD) approaches},
  author={Bakrania, K. and Yates, T. and Rowlands, A. V. and Esliger, D. W. and Bunnewell, S. and Sanders, J. and Davies, M. and Khunti, K. and Edwardson, C. L.},
  journal={PLOS One},
  volume={11},
  number={10},
  pages={e0164045},
  year={2016},
  publisher={Public Library of Science San Francisco, CA USA}
}

@article{bornkamp2018calculating,
  title={Calculating quantiles of noisy distribution functions using local linear regressions},
  author={Bornkamp, B.},
  journal={Computational Statistics},
  volume={33},
  pages={487--501},
  year={2018},
  publisher={Springer}
}

@article{Chinabiobank,
  title={China Kadoorie Biobank of 0.5 million people: survey methods, baseline characteristics and long-term follow-up},
  author={Chen, Z. and Chen, J. and Collins, R. and Guo, Y. and Peto, R. and Wu, F. and Li, L. and {China Kadoorie Biobank (CKB) collaborative group}},
  journal={International Journal of Epidemiology},
  volume={40},
  number={6},
  pages={1652--1666},
  year={2011},
  publisher={Oxford University Press}
}

@article{crainiceanu2012bootstrap,
  title={Bootstrap-based inference on the difference in the means of two correlated functional processes},
  author={Crainiceanu, C. M. and Staicu, A. and Ray, S. and Punjabi, N.},
  journal={Statistics in Medicine},
  volume={31},
  number={26},
  pages={3223--3240},
  year={2012},
  publisher={Wiley Online Library}
}

@article{crainiceanugoldsmithwinbugs,
 title={Bayesian functional data analysis using {WinBUGS}},
 volume={32},
 number={11},
 journal={Journal of Statistical Software},
 author={Crainiceanu, C. M. and Goldsmith, J.},
 year={2010},
 pages={1–33}
}

@book{crainiceanu2024functional,
  title={Functional Data Analysis with R},
  author={Crainiceanu, C. M. and Goldsmith, J. and Leroux, A. and Cui, E.},
  year={2024},
  publisher={Chapman and Hall/CRC}
}

@article{cui2021additive,
  title={Additive functional Cox model},
  author={Cui, E. and Crainiceanu, C. M. and Leroux, A.},
  journal={Journal of Computational and Graphical Statistics},
  volume={30},
  number={3},
  pages={780--793},
  year={2021},
  publisher={Taylor \& Francis}
}

@article{cui2022fast,
  title={Fast univariate inference for longitudinal functional models},
  author={Cui, E. and Leroux, A. and Smirnova, E. and Crainiceanu, C. M.},
  journal={Journal of Computational and Graphical Statistics},
  volume={31},
  number={1},
  pages={219--230},
  year={2022},
  publisher={Taylor \& Francis}
}

@article{cui2022semiparametric,
  title={A semiparametric risk score for physical activity},
  author={Cui, E. and Thompson, E. C. and Carroll, R. J. and Ruppert, D.},
  journal={Statistics in Medicine},
  volume={41},
  number={7},
  pages={1191--1204},
  year={2022},
  publisher={Wiley Online Library}
}

@article{cui2023fast,
  title={Fast Multilevel Functional Principal Component Analysis},
  author={Cui, E. and Li, R. and Crainiceanu, C. M. and Xiao, L.},
  journal={Journal of Computational and Graphical Statistics},
  volume={32},
  number={2},
  pages={366--377},
  year={2023},
  publisher={Taylor \& Francis}
}

@phdthesis{cui2023functional,
  title={Functional Data Analysis Methods for Large Scale Physical Activity Studies},
  author={Cui, E.},
  year={2023},
  school={J.s Hopkins University}
}

@article{di2017patterns,
  title={Patterns of sedentary and active time accumulation are associated with mortality in US adults: The {NHANES} study},
  author={Di, J. and Leroux, A. and Urbanek, J. and Varadhan, R. and Spira, A. P. and Schrack, J. and Zipunnikov, V.},
  journal={BioRxiv},
  pages={182337},
  year={2017},
  publisher={Cold Spring Harbor Laboratory}
}

@article{doherty2017ukbaccel,
    title={Large scale population assessment of physical activity using wrist worn accelerometers: The {UK biobank} study.},
    author={Doherty, A. and Jackson, D. and Hammerla, N. and Plotz, T. and Olivier, P. and Granat, M. H. and White, T. and van Hees, V. T. and Trenell, M. I. and Owen, C. G. and Preece, S. J. and Gillions, R. and Sheard, S. and Peakman, T. and Brage, S. and Wareham, N.},
    journal={PLOS One},
    volume={12},
    number={2},
    pages={e0169649},
    year={2017},
    publisher={Public Library of Science San Francisco, CA USA}
}

@article{eilers1996flexible,
  title={Flexible smoothing with B-splines and penalties},
  author={Eilers, P. H. C. and Marx, B. D.},
  journal={Statistical Science},
  volume={11},
  number={2},
  pages={89--121},
  year={1996},
  publisher={Institute of Mathematical Statistics}
}

@article{gellar2015cox,
  title={Cox regression models with functional covariates for survival data},
  author={Gellar, J. E. and Colantuoni, E. and Needham, D. M. and Crainiceanu, C. M.},
  journal={Statistical Modelling},
  volume={15},
  number={3},
  pages={256--278},
  year={2015},
  publisher={SAGE Publications Sage India: New Delhi, India}
}

@article{ghosal2022scalar,
  title={Scalar on time-by-distribution regression and its application for modelling associations between daily-living physical activity and cognitive functions in Alzheimer’s disease},
  author={Ghosal, R. and Varma, V. R. and Volfson, D. and Urbanek, J. and Hausdorff, J. M. and Watts, A. and Zipunnikov, V.},
  journal={Scientific Reports},
  volume={12},
  number={1},
  pages={11558},
  year={2022},
  publisher={Nature Publishing Group UK London}
}

@article{ghosal2023distributional,
  title={Distributional data analysis via quantile functions and its application to modeling digital biomarkers of gait in Alzheimer’s disease},
  author={Ghosal, R. and Varma, V. R. and Volfson, D. and Hillel, I. and Urbanek, J. and Hausdorff, J. M. and Watts, A. and Zipunnikov, V.},
  journal={Biostatistics},
  volume={24},
  number={3},
  pages={539--561},
  year={2023},
  publisher={Oxford University Press}
}

@article{ghosal2023distributional2,
  title={Distributional outcome regression and its application to modelling continuously monitored heart rate and physical activity},
  author={Ghosal, R. and Ghosh, S. K. and Schrack, J. A and Zipunnikov, V.},
  journal={arXiv preprint arXiv:2301.11399},
  year={2023}
}

@article{goldsmith2011,
  title={Penalized functional regression},
  author={Goldsmith, J. and Bobb, J. and Crainiceanu, C. M. and Caffo, B. and Reich, D.},
  journal={{Journal of Computational and Graphical Statistics}},
  volume={20},
  number={4},
  pages={830--851},
  year={2011},
  publisher={Taylor \& Francis}
}

@Manual{refund,
    title = {refund: Regression with Functional Data},
    author = {Goldsmith, J. and Scheipl, F. and Huang, L. and Wrobel, J. and Di, C. and Gellar, J. and Harezlak, J. and McLean, M. W. and Swihart, B. and Xiao, L. and Crainiceanu, C. M. and Reiss, P. T. and Cui, E.},
    year = {2024},
    note = {R package version 0.1-36},
    url = {https://CRAN.R-project.org/package=refund},
  }

@article{gray1992flexible,
  title={Flexible methods for analyzing survival data using splines, with applications to breast cancer prognosis},
  author={Gray, R. J.},
  journal={Journal of the American Statistical Association},
  volume={87},
  number={420},
  pages={942--951},
  year={1992},
  publisher={Taylor \& Francis}
}

@article{hao2021semiparametric,
  title={Semiparametric inference for the functional Cox model},
  author={Hao, M. and Liu, K. and Xu, W. and Zhao, X.},
  journal={Journal of the American Statistical Association},
  volume={116},
  number={535},
  pages={1319--1329},
  year={2021},
  publisher={Taylor \& Francis}
}

@book{kokoszka2017introduction,
  title={Introduction to functional data analysis},
  author={Kokoszka, P. and Reimherr, M.},
  year={2017},
  publisher={Chapman and Hall/CRC}
}

@article{kong2018flcrm,
  title={FLCRM: Functional linear cox regression model},
  author={Kong, D. and Ibrahim, J. G. and Lee, E. and Zhu, H.},
  journal={Biometrics},
  volume={74},
  number={1},
  pages={109--117},
  year={2018},
  publisher={Wiley Online Library}
}

@article{ledbetter2022cardiovascular,
  title={Cardiovascular mortality risk prediction using objectively measured physical activity phenotypes in NHANES 2003--2006},
  author={Ledbetter, M. K. and Tabacu, L. and Leroux, A. and Crainiceanu, C. M. and Smirnova, E.},
  journal={Preventive medicine},
  volume={164},
  pages={107303},
  year={2022},
  publisher={Elsevier}
}

@article{leroux2019organizing,
  title={Organizing and analyzing the activity data in NHANES},
  author={Leroux, A. and Di, J. and Smirnova, E. and Mcguffey, E. J. and Cao, Q. and Bayatmokhtari, E. and Tabacu, L. and Zipunnikov, V. and Urbanek, J. K and Crainiceanu, C. M.},
  journal={Statistics in biosciences},
  volume={11},
  pages={262--287},
  year={2019},
  publisher={Springer}
}

@phdthesis{leroux2020statistical,
  title={Statistical methods for the analysis of functional data under models with complex association structures},
  author={Leroux, A.},
  year={2020},
  school={J.s Hopkins University}
}

@article{leroux2021quantifying,
  title={Quantifying the predictive performance of objectively measured physical activity on mortality in the {UK Biobank}},
  author={Leroux, A. and Xu, S. and Kundu, P. and Muschelli, J. and Smirnova, E. and Chatterjee, N. and Crainiceanu, C. M.},
  journal={The Journals of Gerontology: Series A},
  volume={76},
  number={8},
  pages={1486--1494},
  year={2021},
  publisher={Oxford University Press US}
}

@article{meng2023quantifying,
  title={Quantifying the Association between Objectively Measured Physical Activity and Multiple Sclerosis in the {UK Biobank}},
  author={Meng, Q. and Cui, E. and Leroux, A. and Mowry, E. M. and Lindquist, M. A. and Crainiceanu, C. M.},
  journal={Medicine and Science in Sports and Exercise},
  volume={55},
  number={12},
  pages={2194-2202},
  year={2023}
}

@article{park2018simple,
  title={Simple fixed-effects inference for complex functional models},
  author={Park, S. Y. and Staicu, A. and Xiao, L. and Crainiceanu, C. M.},
  journal={Biostatistics},
  volume={19},
  number={2},
  pages={137--152},
  year={2018},
  publisher={Oxford University Press}
}

@article{petersen2022modeling,
  title={Modeling probability density functions as data objects},
  author={Petersen, A. and Zhang, C. and Kokoszka, P.},
  journal={Econometrics and Statistics},
  volume={21},
  pages={159--178},
  year={2022},
  publisher={Elsevier}
}

@article{qu2016optimal,
  title={Optimal estimation for the functional cox model},
  author={Qu, S. and Wang, J. and Wang, X.},
  journal={The Annals of Statistics},
  volume={44},
  number={4},
  pages={1708--1738},
  year={2016}
}

@book{ramsay2005fitting,
  title={Functional Data Analysis},
  author={Ramsay, J. O. and Silverman, B. W.},
  year={2005},
  publisher={Springer}
}

@book{ruppert2003semiparametric,
  title={Semiparametric Regression},
  author={Ruppert, D. and Wand, M. P. and Carroll, R. J.},
  number={12},
  year={2003},
  publisher={Cambridge University Press}
}

@article{sergazinov2023case,
  title={A case study of glucose levels during sleep using multilevel fast function on scalar regression inference},
  author={Sergazinov, R. and Leroux, A. and Cui, E. and Crainiceanu, C. M. and Aurora, R. N. and Punjabi, N. M. and Gaynanova, I.},
  journal={Biometrics},
  volume={79},
  number={4},
  pages={3873--3882},
  year={2023},
  publisher={Oxford University Press}
}

@article{smirnova2020predictive,
  title={The predictive performance of objective measures of physical activity derived from accelerometry data for 5-year all-cause mortality in older adults: National Health and Nutritional Examination Survey 2003--2006},
  author={Smirnova, E. and Leroux, A. and Cao, Q. and Tabacu, L. and Zipunnikov, V. and Crainiceanu, C. M. and Urbanek, J. K},
  journal={The Journals of Gerontology: Series A},
  volume={75},
  number={9},
  pages={1779--1785},
  year={2020},
  publisher={Oxford University Press US}
}

@article{tabacu2020,
  title={Quantifying the Varying Predictive Value of Physical Activity Measures Obtained from Wearable Accelerometers on All-Cause Mortality over Short to Medium Time Horizons in NHANES 2003-2006},
  author={Tabacu, L. and Ledbetter, M. and Leroux, A. and Crainiceanu, C.M. and Smirnova, E.},
  journal={Sensors},
  volume={21},
  number={1},
  pages={1779--1785},
  year={2020}
}

@article{allofus2019,
  title={The “All of Us” Research Program},
  author={{The All of Us Research Program Investigators}},
  journal={New England Journal of Medicine},
  volume={381},
  pages={668-676},
  year={2019}
}

@article{van2013separating,
  title={Separating movement and gR.ty components in an acceleration signal and implications for the assessment of human daily physical activity},
  author={Van Hees, V. T. and Gorzelniak, L. and Dean Le{\'o}n, E. C. and Eder, M. A. and Pias, M. and Taherian, S. and Ekelund, U. and Renstr{\"o}m, F. and Franks, P. W. and Horsch, A. and others},
  journal={PLOS One},
  volume={8},
  number={4},
  pages={e61691},
  year={2013},
  publisher={Public Library of Science San Francisco, USA}
}

@article{verweij1994penalized,
  title={Penalized likelihood in Cox regression},
  author={Verweij, P. J. M. and Van Houwelingen, H. C.},
  journal={Statistics in Medicine},
  volume={13},
  number={23-24},
  pages={2427--2436},
  year={1994},
  publisher={Wiley Online Library}
}

@book{wood2017generalized,
  title={Generalized Additive Models: an Introduction with R},
  author={Wood, S. N.},
  year={2017},
  publisher={CRC Press}
}

@article{xiao2016,
    author = {Xiao, L. and  Zipunnikov, V. and Ruppert, D. and Crainiceanu, C.M.},
    title = "{Fast covariance estimation for high-dimensional functional data}",
    journal = {Statistics and Computing},
    volume = {26},
    number = {1},
    pages = {409–421},
    year = {2016}
}

@article{zhao2023evaluating,
  title={Evaluating the prediction performance of objective physical activity measures for incident Parkinson’s disease in the UK Biobank},
  author={Zhao, A. and Cui, E. and Leroux, A. and Lindquist, M. A. and Crainiceanu, C. M.},
  journal={Journal of Neurology},
  volume={270},
  number={12},
  pages={5913--5923},
  year={2023},
  publisher={Springer}
}

\end{document}